# Nonlinearity Characteristic of High Impedance Fault at Resonant Distribution Networks: Theoretical Basis to Identify the Faulty Feeder

Mingjie Wei, Hengxu Zhang, Fang Shi, Weijiang Chen, *Senior Member, IEEE,* Vladmir Terzija, *Fellow, IEEE*

*Abstract*— Feeder identification is indispensable for distribution networks to locate faults at a specific feeder, especially when measuring devices are insufficient to locate faults more precisely. For the high impedance fault (HIF), the feeder identification is much more challengeable and related approaches are still in the early stage. This paper thoroughly reveals the nonlinearity characteristics of different feeders when a HIF happens at a three-wire system that is with the resonant grounded neutral (RGN). Firstly, the diversity of nonlinearity existing in HIFs is explained from the perspective of energy. Then, the nonlinearity of zero-sequence current that differs between healthy and faulty feeders are deduced theoretically. Effects of the detuning index of Petersen coil, the damping ratio of system, and the length of feeder are all considered. Afterward, these theoretical conclusions are verified by the HIF cases experimented in a real 10kV system. Finally, after indicating the problems of a classic phase-relationship-based algorithm, we suggest an improved method based on the phase differences between the harmonic currents of different feeders. The effectiveness of the method has been verified.

## I. INTRODUCTION

HIGH impedance fault (HIF) is a common fault type at medium-voltage (MV) distribution networks [1]. Generally, HIFs are the single-line-to-ground (SLG) faults, happening in overhead lines when the line conductor breaks off or sags to touch the high impedance grounding materials, like soil, concrete, asphalt, grass, and tree limb. The current amplitudes of HIFs are generally lower than 50A, or even below 1 A for some grounding materials with extremely high impedances [2]. In these cases, traditional overcurrent relays, which are still widely applied at networks nowadays, are unable to guarantee reliability. According to the early systematical staged tests carried out by Texas A&M University, only 17.5% of 200 HIFs can be detected by traditional overcurrent relays [3]. When the faulted line establishes electrical conduction with the ground, the ignition of air arc and the breakdown of solid dielectric always exist [4], releasing energies and causing nonlinearities. The long-time existence of HIFs could raise the risks of fire and shock accidents. The severe fire hazards recently occurring in Australia, the United States, and Brazil, are confirmed to result from HIFs initially [5]. It further emphasizes the necessity to detect and isolate HIFs.

Faulty feeder identification, which is a part of fault location, is still important for nowadays distribution systems. At a practical network, a substation is commonly connected by several feeders, which further extend to a number of branches and supply for an area of dozens of kilometers. In order to locate the fault within a small area or even to a specific point, most techniques have to rely on the advanced devices which are deployed all over each feeder. However, the limitation of investment cannot always support that high-density equipment deployment at distribution networks. As a result, the identification of faulty feeder is still economical and practical, which only needs the devices at the substation and the start of feeders. After the faulty feeder is identified, maintainers can fast patrol the specific transmission line for the fault position or switch off the whole feeder if necessary.

The feeder identification is commonly triggered by the successful detection of a HIF, where the latter technique has been researched over 40 years and achieved inspiring progress [6]-[8]. An approach to identify the faulty feeder is significantly dependent on the neutral type and the topology of a network. From a worldwide perspective, the distribution network is classified as four-wire and three-wire systems. The four-wire system exists in the United States, Canada, Brazil, Mexico, and Australia, etc., where the neutral of the network is solidly grounded and a neutral line spreads over the network. For the three-wire system that is applied in many other countries like China, Europe, and Japan [9],[10], isolated neutral (IN), resonant grounded neutral (RGN), and low-resistor grounded neutral (LRGN) are selectively applied on different conditions. Although the four-wire system possesses some advantages like the economy, it faces larger problems in HIF diagnosis due to the interference of load current [11]. In the three-wire system, the zero-sequence current (ZSC) is close to zero at normal states and unaffected by (unbalanced) currents of loads, due to the isolation of load transformers [12]-[14]. In this way, fault features will not be submerged, and the reliability to identify the fault feeders can be significantly improved. In this paper, only the three-wire system is discussed.

Traditionally, fundamental phasors are the easiest way to isolate faults. Fig.1 (a)~(c) presents the phasor diagram of fundamental ZSCs and zero-sequence voltage (ZSV) at the three-wire system with three common neutrals, respectively. For the IN (Fig.1a) and LRGN (Fig.1c) networks, faulty and healthy feeders can be easily distinguished just by fundamental phases and amplitudes. Based

This work was supported by the National Key R&D Program of China (2017YFB0902800) and the Science and Technology Project of State Grid Corporation of China (52094017003D).

M. Wei, F. Shi and H. Zhang are all with the Key Laboratory of Power System Intelligent Dispatch and Control Ministry of Education, Shandong University, Jinan, 250061, China (e-mail: zhanghx@sdu.edu.cn).
W. Chen is with the State Grid Corporation of China (e-mail:weijiang-chen@sgcc.cn).
V. Terzija is with the School of Electrical and Electronic Engineering, The University of Manchester, U.K. (e-mail: vladimir.terzija@manchester.ac.uk).

on this characteristic, algorithms in [14] and [15] are proposed to improve anti-noise ability further. However, at the RGN network (Fig.1b), the faulty feeder cannot be picked out only using phasors. For the low impedance fault (LIF), this problem is solved by utilizing the direction of high-frequency transients [16],[17], which has been applied over 15 years in China [18]. However, for HIFs, the high-frequency transients are gradually suppressed as the increase of fault impedance [9]. To solve this problem, a rule-based algorithm is proposed in [9] and a pattern-recognition-based algorithm is proposed in [12]. They identify the HIF feeders with low-frequency transient projections and waveform distortions, respectively. However, their sensitivities are restricted when the detuning of the Petersen coil and nonlinear distortion vary.

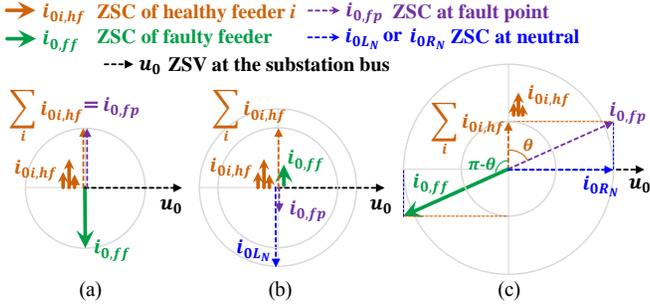

Fig.1 Phasors of fundamental ZSCs at different feeders: (a) IN network; (b) RGN network; (c) LRGN network.

An important reason that the reliability to identify HIF feeder is not high because most algorithms are based on the 'phenomenon' of some real-world cases or based on incomplete modelings. As a result, the algorithms cannot apply to all fault conditions. The contributions of this paper are summarized as: 1) the diversity of HIF nonlinearity is explained from the aspect of energy balance, clarifying how the nonlinearity of fault current is affected by external environments; 2) the nonlinearity characteristics of healthy and faulty feeders at RGN networks are theoretically deduced; 3) the detuning index of Petersen coil, the damping ratio of the system, and the length of feeder, etc., are all considered in the theory. The theory is verified by the real-world HIFs experimented in a 10kV system. 4) Based on the theory, this paper provides an improved 3rd-harmonic-based algorithm to identify the faulty feeder. Illustrations and the comparisons with a classic 3rd-harmonic-based method show the effectiveness and significant improvements.

The rest of the paper is organized as follows: Section II introduces the nonlinearity of HIF and how it is affected by the external environment. Section III analyzes the differences of nonlinearity characteristics between different feeders from the theoretical aspect. Considering the phase-to-earth resistances in the practical resonant system, Section IV also takes the damping ratio into account, so as to modify the theoretical analysis. In Section V, an improved phase-relationship-based method is proposed. Effectiveness and advantages are verified. Finally, conclusions are drawn in Section VI.

## II. FEATURES OF HIF NONLINEARITY

### A. Analysis of Distortion Features by Energy

The nonlinearity of HIF is generally caused by the arcing process. For the arc column, the stored energy $Q$ can be expressed according to the energy balance equation [19]:

$$\frac{dQ}{dt} = u \cdot i - P_{Loss} \quad (1)$$

where, $P_{Loss}$ represents the power loss to the air, and $u$, $i$ represent the arc voltage and current, respectively. As the arc is nearly resistive, equation (1) can be further transformed as:

$$\frac{dQ}{dg_{arc}} \cdot \frac{dg_{arc}}{dt} = \frac{dQ}{dg_{arc}} \cdot \frac{-g_{arc}^2 \cdot dR_{arc}}{dt} = u \cdot i - P_{Loss} \quad (2)$$

where, $g_{arc}$ is the arc conductance and $R_{arc}$ is the arc resistance ($R_{arc}=1/g_{arc}$). Then, (2) can be written as:

$$\frac{1}{R_{arc}} \frac{dR_{arc}}{dt} = \frac{P_{Loss} - u \cdot i}{g_{arc} \cdot dQ/dg_{arc}} = \frac{1}{\tau}(P_{Loss} - u \cdot i) \quad (3)$$

where, $\tau = g_{arc} \cdot dQ/dg_{arc}$. Then, $R_{arc}$ is expressed as:

$$R_{arc} = e^{\int \frac{1}{\tau}(P_{Loss} - u \cdot i)dt} \quad (4)$$

Arcing nonlinearity is essentially caused by the nonlinearity of $R_{arc}$, which is affected by the energy conversion in the arc gap. Because the input power $u \cdot i$ is the dependent variable, $R_{arc}$ is only affected by $\tau$ and $P_{Loss}$. Herein, $P_{Loss}$ reflects the energy dissipation ability in the arc column, which is closely related to the movability and thermal conductivity of air space and ground medium. About $\tau$, it reflects how fast arc resistance changes as the stored energy.

Simulations are conducted to figure out how $R_{arc}$ is affected by the two parameters. Some examples are shown in Fig.2, where $P_{Loss}$ and $\tau$ are assumed as constants for simplification, just like the Mayr model. When modeling the HIF, a large linear resistor $R_T$ is usually in series with the nonlinear $R_{arc}$, so as to decrease the RMS of current.

$R_{arc}$ exhibits as a 'spike' curve twice in each cycle and reaches the maximum when $dR_{arc}/dt=0$, i.e., $u \cdot i = P_{Loss}$. Because $u \cdot i = 0$ at the zero-crossing of current, the maximal $R_{arc}$ will show increasing deviation relative to zero-crossings as $P_{Loss}$ increases, like from Fig.2(a) to (c). Besides, $P_{Loss}$ and $\tau$ respectively determines how long and how fast $R_{arc}$ increases, which together affect the level of $R_{arc}$.

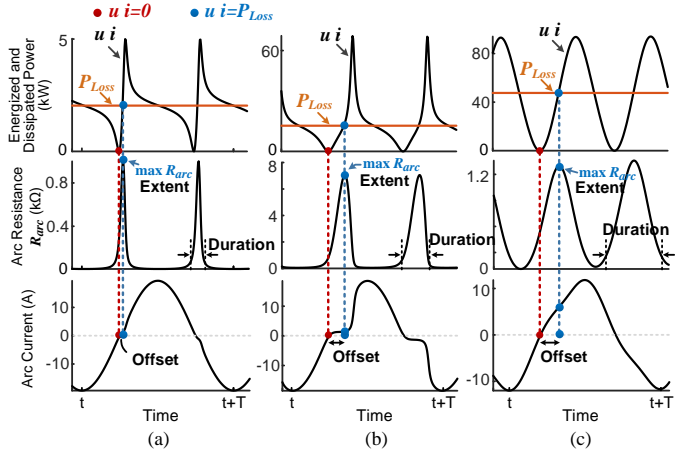

Fig.2 Effects of $P_{Loss}$ (kW) and $\tau$ (kW·s) on the waveform distortion: (a) $P_{Loss}=2$, $\tau=0.3$; (b) $P_{Loss}=16$, $\tau=1.67$; (c) $P_{Loss}=46$, $\tau=3.3$.

In our previous work [19], the relationship between $R_{arc}$ and distortion has been discussed. In brief, the distortion of HIF current can be described from three aspects: offset, duration, and extent. As shown in Fig.2, they respectively represent what time does the distortion happen, how long does the observable distortion exist, and how severe the distortion is. Specifically, the distortion offset is closely related to the deviation of maximal $R_{arc}$, i.e., the deviation of $u \cdot i = P_{Loss}$ from $u \cdot i = 0$. The distortion duration is determined by the duration when $R_{arc}$ reaches a certain high value, and the distortion extent is reflected by the value of maximal $R_{arc}$.

In addition, Fig.3 presents the 2$^{nd}$~7$^{th}$ harmonic components of the HIF current in Fig.2 (a)~(c), respectively. It verifies that odd harmonics dominate the nonlinearity of HIF current. Meanwhile, the amplitudes and phases of different harmonics are all affected when the distortion of current varies as the external environment.

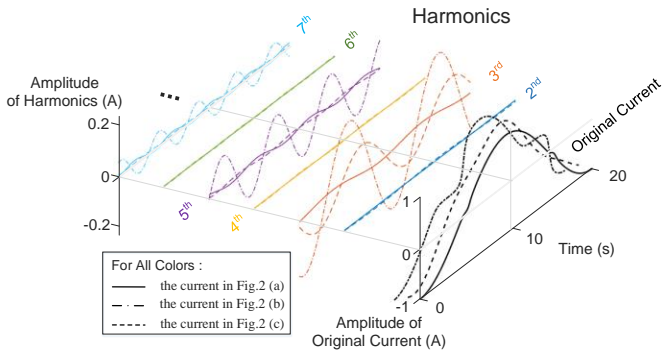

Fig.3 Curves of different harmonic components when the distortion of current varies

In conclusion, the external environment of HIF, like material, humidity, and structure of the ground surface, affects the energy conversion of arc. Then, the energy conversion, reflected by $P_{Loss}$ and $\tau$, controls the curve of $R_{arc}$. Finally, $R_{arc}$ determines the nonlinearity characteristic of fault current, which reflects on the distortions and harmonics. For this reason, the diversity of nonlinearity, as well as its effects on a HIF diagnosis method, must be considered. A more universal characteristic quantity for the feeder identification is needed, to guarantee reliability under as many conditions as possible.

### B. Artificial Experiments in A Real 10kV System

Artificial HIFs have been tested at a real 10kV 50Hz distribution network, the topology of which is shown in Fig.4. In this paper, these real HIFs are used to explain the practical fault features and verify the theories.

The neutral of RGN network is grounded with a Petersen coil. There are four feeders under operations, with the combination of underground cables and overhead lines, as shown in Fig.4. Specific lengths of feeders are also labeled in the figure. For each feeder, the load current is around 15~20A. HIFs are tested after grounding the energized lines to different surfaces, including soil, cement, asphalt, and grass, etc. The surface humidity is simply classified as wet and dry. Signals are measured by devices with 6.4 kHz sampling frequency, which are deployed at the start of each feeder (M1~M4), outside the substation (M5), and at the fault point (Mf).

In Fig.5, black curves are the fault currents of real-world HIFs that happened in different scenarios. The diversity of waveform distortion confirms the discussions in the previous section. In the practical application of HIF diagnosis, this distortion diversity has to be considered, especially for the slight distortions like Fig.5 (d)-(e), or large distortion offsets in Fig.5 (f)-(g). They are more likely to be overlooked by traditional methods, leading to mistakes.

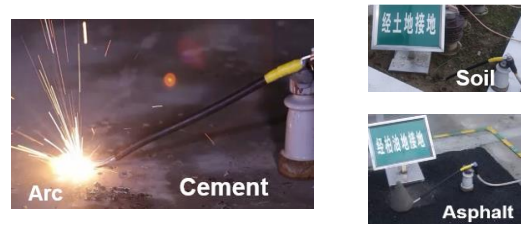

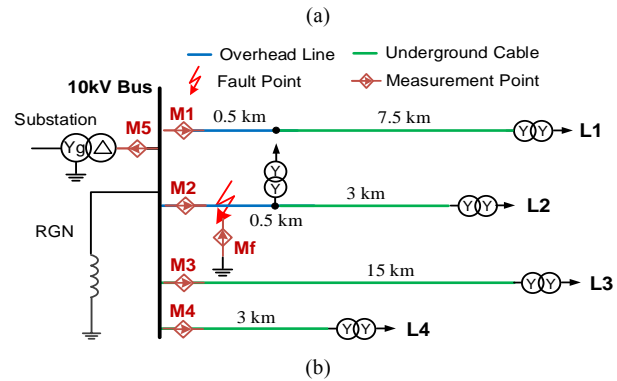

Fig.4 Experiments of HIFs in a real 10 kV distribution network. (a) Photographs of HIF experiments. (b) Topology of the 10kV distribution system.

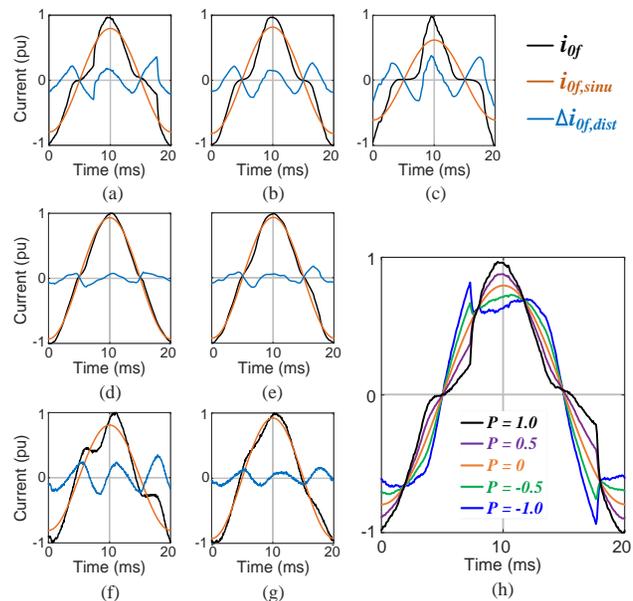

Fig.5 Sinusoidal and distortional components of HIF current experimented in a real 10kV distribution network: (a) wet asphalt concrete; (b) dry soil; (c) wet soil; (d) wet cement; (f) wet asphalt concrete; (f) dry grass; (g) wet grass; (h) the current waveforms when the distortional content varies.

## III. DISTORTION DIFFERENCES BETWEEN FEEDERS

For the RGN network, distortion differences of ZSCs at healthy and faulty feeders are theoretically deduced in this section.

### A. Sinusoidal Component and Distortional Component

As is well-known, any signal can be resolved as the linear superposition of sine waves with different frequencies. Therefore, the ZSC at the fault point, denoted as $i_{0f}$, is written as:

$$i_{0f} = i_{0f,sinu} + \Delta i_{0f,dist}$$
$$\begin{cases} i_{0f,sinu} = A_0 \sin(\omega_0 t + \varphi_0) \\ \Delta i_{0f,dist} \approx \sum_{k=2}^{m} \Delta i_{0f,dist}^{(H,k)} = \sum_{k=2}^{m} A_{H,k} \sin(k\omega_0 t + \varphi_{H,k}) \end{cases} \quad (5)$$

where the sinusoidal component and distortional component of $i_{0f}$ are defined as $i_{0f,sinu}$ and $\Delta i_{0f,dist}$, respectively. $\omega_0$, $\varphi_0$, and $A_0$ are the fundamental radian frequency, phase (angle), and peak amplitude, respectively. $\Delta i_{0f,dist}$ is combined with various frequencies of harmonics, where the $\varphi_{H,k}$ and $A_{H,k}$ represent the phase and peak amplitude of the kth harmonic distortional component (denoted as $\Delta i_{0f,dist}^{(H,k)}$), respectively. The white noise, impulse noise, and inter-harmonics are neglected as distortions are mainly caused by the low-order integer harmonics.

In Fig.5, the $i_{0f}$ of some real HIFs experimented in the aforementioned 10kV system are presented, together with their two components. Specifically, $i_{0f,sinu}$ is achieved by FFT while $\Delta i_{0f,dist}$ is achieved by $i_{0f} - i_{0f,sinu}$. Take Fig.5 (a) as an example and change the content of $\Delta i_{0f,dist}$ as:

$$i_{0X} = i_{0f,sinu} + P \cdot \Delta i_{0f,dist} \quad (6)$$

Respectively set $P$ as -1, -0.5, 0, 0.5, 1, and the corresponding $i_{0X}$ are shown in Fig.5 (h). Define the conditions when $P > 0$ as 'positive superposition', and as 'negative superposition' when $P < 0$. As a result, $i_{0f}$ is always with 'positive superposition'. The currents with different superpositions will present significantly different distortion and the opposite phases of harmonic components.

### B. Equivalent Zero-Sequence Circuit

An unbalanced three-wire system can be transformed into a modulus system by Karrenbauer transformation [20]. The transformed system is combined with three mode networks that are self-balanced, including 1 MODE, 2 MODE, and 0 MODE networks. The 0 MODE network is exactly the so-called zero-sequence network. An RGN system with n feeders (L1~Ln, like the 4-feeder system in Fig.4) can be transformed into the modulus system in Fig.6(a), where the HIF is assumed to happen at feeder $n$. The impedances of feeders are all simplified as the Γ-type circuits. The variables in Fig.6, as well as in the following Fig.7, are defined in TABLE 1.

TABLE 1 VARIABLES IN THE EQUIVALENT CIRCUIT

| | |
|---|---|
| $Z_{pi}$ | Impedance of feeder $i$ ($i$=1,2,..,n-1) in the $p$ ($p$=1,2,0) MODE network |
| $C_{pi}$ | Phase-to-earth capacitance of feeder $i$ ($i$=1,2,..,n-1) in the $p$ ($p$=1,2,0) MODE network |
| $Z_{pn,bef}$ | The impedance of the faulty feeder (feeder $n$) before the fault point, in the $p$ ($p$=1,2,0) MODE network. |
| $Z_{pn,aft}$ | The impedance of the faulty feeder (feeder $n$) after the fault point, in the $p$ ($p$=1,2,0) MODE network. |
| $C_{pn,bef}$ | The phase-to-earth capacitance of the faulty feeder (feeder $n$) before the fault point, in the $p$ ($p$=1,2,0) MODE network. |
| $C_{pn,aft}$ | The phase-to-earth capacitance of the faulty feeder (feeder $n$) after the fault point, in the $p$ ($p$=1,2,0) MODE network. |
| $L$ | Equivalent zero-sequence (0 MODE) inductance of the Petersen coil (equal to 3 times the coil inductance) |
| $R_{HIF}$ | Equivalent zero-sequence (0 MODE) resistance of fault (3 times the grounding resistance) |
| $u_f$ | Equivalent virtual voltage source, which has the same amplitude with and opposite phase to the pre-fault phase-to-ground voltage at the fault point |
| $u_p$ | $p$ ($p$=1,2,0) MODE voltage at the fault point |
| $u_{0b}$ | Zero-sequence (0 MODE) voltage at the substation bus |
| $i_{0f}$ | Zero-sequence (0 MODE) current at the fault point |
| $i_{0i}$ | Zero-sequence (0 MODE) current at the beginning of feeder $i$ |
| $i_{0L}$ | Zero-sequence (0 MODE) current through the Petersen coil |
| $i_{0C_{0i}}$ | Zero-sequence (0 MODE) current of the phase-to-earth capacitance at feeder $i$ |
| $Z_1, Z_2$ | The total impedances of 1 MODE and 2 MODE networks |

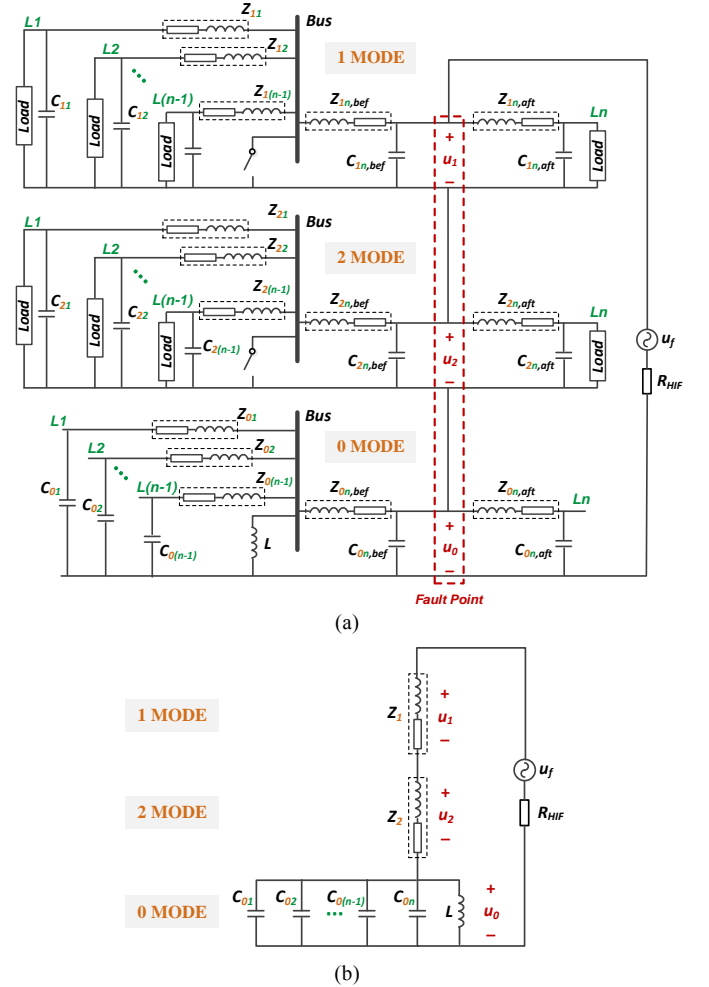

Fig.6 The modulus system transformed by Karrenbauer transformation: (a) without simplification of line impedances, (b) with the simplification of line impedances.

For the three-wire system, load impedances only exist in 1 MODE and 2 MODE networks, but not in the 0 MODE network due to the isolation of delta-wye load transformers [9], [14], [21]. As this paper mainly focuses on the fundamental component and the low-order harmonics of signals, the branches of phase-to-earth capacitances in the 1 MODE and 2 MODE networks ($C_{1i}$, $C_{2i}$, $C_{1n,bef}$, $C_{1n,aft}$, $C_{2n,bef}$, and $C_{2n,aft}$) can be simplified as open-circuited, while the line impedances of 0 MODE ($Z_{0i}$, $Z_{0n,bef}$, and $Z_{0n,aft}$) can be regarded as short-circuited. Then, the modulus system in Fig.6(a) can be transformed as Fig.6(b). According to the Thevenin theorem, the external voltage source of 0 MODE network is $u_f$ and the external equivalent impedance $Z=R_{HIF}+Z_1+Z_2$. As a result, the 0 MODE (zero-sequence) network can be presented as Fig.7, where $u_{0b}$ is the measured zero-sequence voltage at the start of feeder and $u_{0b} \approx u_0$.

In Fig.7, the zero-sequence phase-to-earth capacitance $C_{0i}$ is affected by the length of feeder, the type of feeder (overhead line or underground cable), and the unbalance of three phases. In addition to the above factors, $Z$ is also affected by loads and the unbalance of self/shunt line impedances.

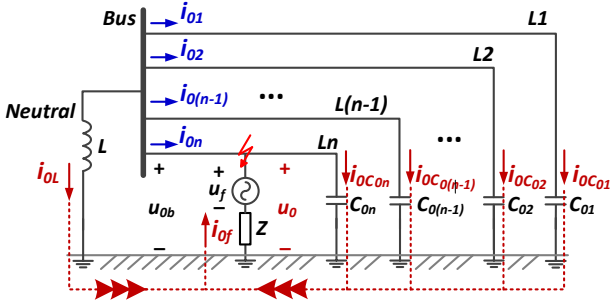

Fig.7 Equivalent zero-sequence circuit of an *n*-feeder resonant network.

### C. Differences Between Feeders

*1) Expressions of Sinusoidal and Distortional Components:*

Set a HIF in feeder $n$, then $i_{0f}$ can be expressed as:

$$i_{0f}=i_{0L}+\sum_{i=1}^{n} i_{0C_{0i}} \quad (7)$$

Then, $i_{0L}$ and $i_{0C_{0i}}$ are also respectively resolved into a sinusoidal component and a distortional component:

$$i_{0f}=i_{0L,sinu}+\sum_{k=2}^{m}\Delta i_{0L,dist}^{(H,k)}+\sum_{i=1}^{n}\left(i_{0C_{0i},sinu}+\sum_{k=2}^{m}\Delta i_{0C_{0i},dist}^{(H,k)}\right) \quad (8)$$

where the sinusoidal and distortional components of $i_{0L}$ and $i_{0C_{0i}}$ are accordingly expressed. Further, according to Fig.7:

$$u_{0b}=\frac{1}{C_{0i}}\int i_{0C_{0i}}dt=\frac{1}{C_{0\Sigma}}\int \sum_{i=1}^{n} i_{0C_{0i}} dt = L\frac{di_{0L}}{dt}$$

$$\Rightarrow u_{0b} = u_{0b,sinu} + \sum_{k=2}^{m}\Delta u_{0b,dist}^{(H,k)}$$

$$=\frac{1}{C_{0i}}\int \left(i_{0C_{0i},sinu}+\sum_{k=2}^{m}\Delta i_{0C_{0i},dist}^{(H,k)}\right)dt \quad (9)$$

$$=\frac{1}{C_{0\Sigma}}\int \sum_{i=1}^{n}\left(i_{0C_{0i},sinu}+\sum_{k=2}^{m}\Delta i_{0C_{0i},dist}^{(H,k)}\right)dt$$

$$=L\frac{d\left(i_{0L,sinu}+\sum_{k=2}^{m}\Delta i_{0L,dist}^{(H,k)}\right)}{dt}$$

where, $C_{0\Sigma}=\sum_{i=1}^{n} C_{0i}$.

It is known that the sine waves with different frequencies are mutually orthogonal. It means the sinusoidal and distortional components, as well as different orders of harmonics, cannot transform into each other. As a result, based on (5)-(9), we have:

$$\begin{cases} i_{0f,sinu}=i_{0L,sinu}+\sum_{i=1}^{n} i_{0C_{0i},sinu} \\ u_{0b,sinu}=\frac{\int i_{0C_{0i},sinu}dt}{C_{0i}}=\frac{\int \sum_{i=1}^{n} i_{0C_{0i},sinu} dt}{C_{0\Sigma}}=L\frac{di_{0L,sinu}}{dt} \end{cases} \quad (10)$$

$$\begin{cases} \Delta i_{0f,dist}^{(H,k)}=\Delta i_{0L,dist}^{(H,k)}+\sum_{i=1}^{n}\Delta i_{0C_{0i},dist}^{(H,k)} \\ \Delta u_{0b,dist}^{(H,k)}=\frac{\int \Delta i_{0C_{0i},dist}^{(H,k)}dt}{C_{0i}}=\frac{\int \sum_{i=1}^{n}\Delta i_{0C_{0i},dist}^{(H,k)} dt}{C_{0\Sigma}}=L\frac{d\Delta i_{0L,dist}^{(H,k)}}{dt} \end{cases} \quad (11)$$

It means the superposition principle in linear systems is still valid for the sinusoidal and distortional components.

For $i_{0f}$, when the phase of its sinusoidal component $i_{0f,sinu}$ is $\varphi_0$ as defined in (5), the phases of $i_{0L,sinu}$ and $i_{0C_{0i},sinu}$ should respectively be $\varphi_0$ and $\varphi_0+\pi$ according to (10). For $\Delta i_{0f,dist}^{(H,k)}$, according to (5) and (11), it can be expressed as:

$$\Delta i_{0f,dist}^{(H,k)} = \Delta i_{0L,dist}^{(H,k)} + \sum_{i=1}^{n} \Delta i_{0C_{0i},dist}^{(H,k)}$$

$$= \Delta i_{0L,dist}^{(H,k)} + LC_{0\Sigma}\frac{d^2\Delta i_{0L,dist}^{(H,k)}}{dt^2} = A_{H,k}\sin(k\omega_0 t+\varphi_{H,k}) \quad (12)$$

This second-order non-homogeneous linear (NHL) equation can be solved as:

$$\begin{cases} \Delta i_{0L,dist}^{(H,k)} = \frac{1}{1-k^2\omega_0^2 LC_{0\Sigma}} A_{H,k}\sin(k\omega_0 t+\varphi_{H,k}) \\ \Delta i_{0C_{0i},dist}^{(H,k)} = \frac{-k^2\omega_0^2 LC_{0i}}{1-k^2\omega_0^2 LC_{0\Sigma}} A_{H,k}\sin(k\omega_0 t+\varphi_{H,k}) \end{cases} \quad (13)$$

*2) Expressions of ZSCs at Different Feeders:*

For an RGN network, the detuning index $v$ reflects the ability of the Petersen coil to compensate for the capacitive currents of the network [9], and it is expressed as:

$$v = \frac{\sum_{i=1}^{n} A_{MC_{0i}} - A_{ML}}{\sum_{i=1}^{n} A_{MC_{0i}}} = 1 - \frac{1}{\omega_0^2 LC_{0\Sigma}} = \frac{\omega_0^2 LC_{0\Sigma}-1}{\omega_0^2 LC_{0\Sigma}} \quad (14)$$

where, $A_{MC_{0i}}$ and $A_{ML}$ represent the peak amplitude of $i_{0C_{0i},sinu}$ and $i_{0L,sinu}$, respectively. Ideally, $v\in[-0.1, 0)$ so that $\omega_0^2 LC_{0\Sigma}\in[0.9091, 1)$. Then, with (14), ratios of the following peak amplitudes can be expressed:

$$\begin{cases} \frac{A_{ML}-\sum_{i=1}^{n-1} A_{MC_{0i}}}{A_{ML}-\sum_{i=1}^{n} A_{MC_{0i}}}=\frac{\omega_0^2 L(C_{0\Sigma}-C_{0n})-1}{\omega_0^2 LC_{0\Sigma}-1}=1-\frac{C_{0n}}{vC_{0\Sigma}} \\ \frac{A_{MC_{0i}}}{A_{ML}-\sum_{i=1}^{n} A_{MC_{0i}}}=\frac{\omega_0^2 LC_{0i}}{1-\omega_0^2 LC_{0\Sigma}}=\frac{C_{0i}}{-vC_{0\Sigma}} \\ \frac{A_{ML}}{A_{ML}-\sum_{i=1}^{n} A_{MC_{0i}}}=\frac{1}{1-\omega_0^2 LC_{0\Sigma}}=1-\frac{1}{v} \end{cases} \quad (15)$$

When $i_{0f}$ is expressed as:

$$i_{0f} = i_{0L} + \sum_{i=1}^{n} i_{0C_{0i}} = i_{0f,sinu} + \Delta i_{0f,dist}$$

$$= \left(A_{ML}-\sum_{i=1}^{n} A_{MC_{0i}}\right)\sin(\omega_0 t+\varphi_0) + \sum_{k=2}^{m} A_{H,k}\sin(k\omega_0 t+\varphi_{H,k}) \quad (16)$$

the ZSCs at the start of the faulty feeder ($i_{0n}$), healthy feeders ($i_{0i(i\neq n)}$), and substation feeder ($i_{0L}$), can be respectively expressed as (17)-(19). During the deducing of (17)-(19), the equations (14) and (15) are used for transformation. For the sake of brevity, we only present the process of (17).

$$\begin{aligned} i_{0n} &= -i_{0f} + i_{0C_{0n}} = -\left(i_{0L} + \sum_{i=1}^{n-1} i_{0C_{0i}}\right) = -\left(i_{0L,sinu} + \sum_{i=1}^{n-1} i_{0C_{0i},sinu}\right) - \sum_{k=2}^{m}\left(\Delta i_{0L,dist}^{(H,k)} + \sum_{i=1}^{n-1} \Delta i_{0C_{0i},dist}^{(H,k)}\right) \\ &= -\left(A_{ML} - \sum_{i=1}^{n-1} A_{MC_{0i}}\right)\sin(\omega_0 t + \varphi_0) - \sum_{k=2}^{m}\left[1 + \frac{k^2\omega_0^2 LC_{0n}}{1 - k^2\omega_0^2 LC_{0\Sigma}}\right] A_{H,k}\sin(k\omega_0 t + \varphi_{H,k}) \\ &= \frac{\omega_0^2 L(C_{0\Sigma} - C_{0n}) - 1}{\omega_0^2 LC_{0\Sigma} - 1}(-i_{0f}) - \sum_{k=2}^{m} \frac{\omega_0^2 LC_{0n}(1-k^2)}{(\omega_0^2 LC_{0\Sigma} - 1)(1 - k^2\omega_0^2 LC_{0\Sigma})} A_{H,k}\sin(k\omega_0 t + \varphi_{H,k}) \\ &= \left(\frac{1}{-v}\right)\left[\left(-v + \frac{C_{0n}}{C_{0\Sigma}}\right)(-i_{0f}) + \sum_{k=2}^{m} \frac{C_{0n}}{C_{0\Sigma}} \frac{1-k^2}{1 - k^2\omega_0^2 LC_{0\Sigma}} A_{H,k}\sin(k\omega_0 t + \varphi_{H,k})\right] \end{aligned} \tag{17}$$

$$i_{0i(i\neq n)} = i_{0C_{0i}} = i_{0C_{0i},sinu} + \sum_{k=2}^{m} \Delta i_{0C_{0i},dist}^{(H,k)} = \frac{C_{0i}}{-vC_{0\Sigma}}\left[(-i_{0f}) + \sum_{k=2}^{m} \frac{1-k^2}{1 - k^2\omega_0^2 LC_{0\Sigma}} A_{H,k}\sin(k\omega_0 t + \varphi_{H,k})\right] \tag{18}$$

$$i_{0L} = i_{0L,sinu} + \sum_{k=2}^{m} \Delta i_{0L,dist}^{(H,k)} = \left(\frac{1}{-v}\right)\left[(1-v)i_{0f} - \sum_{k=2}^{m} \frac{1-k^2}{1 - k^2\omega_0^2 LC_{0\Sigma}} A_{H,k}\sin(k\omega_0 t + \varphi_{H,k})\right] \tag{19}$$

In the final expressions of (17)~(19), when $k \geq 2$, it has:

$$\frac{1-k^2}{1 - k^2\omega_0^2 LC_{0\Sigma}} \in \left(\frac{1}{\omega_0^2 LC_{0\Sigma}}, \frac{3}{4\omega_0^2 LC_{0\Sigma} - 1}\right] \tag{20}$$

After $\omega_0^2 LC_{0\Sigma}$ is determined as a value in [0.9091, 1), the widest range of $\frac{1-k^2}{1 - k^2\omega_0^2 LC_{0\Sigma}}$ is (1.100, 1.138]. Simplify $\frac{1-k^2}{1 - k^2\omega_0^2 LC_{0\Sigma}}$ as a constant, then (17)~(19) can be rewritten as (21)~(23), respectively:

$$\begin{aligned} i_{0n} &= \left(\frac{1}{-v}\right)[P_1(-i_{0f}) + P_2\Delta i_{0f,dist}] \\ &= -\left(\frac{1}{-v}\right)P_1\left[i_{0f,sinu} + \frac{P_1 - P_2}{P_1}\Delta i_{0f,dist}\right] \end{aligned} \tag{21}$$

$$\begin{aligned} i_{0i(i\neq n)} &= \left(\frac{1}{-v}\right)[P_3(-i_{0f}) + P_4\Delta i_{0f,dist}] \\ &= -\left(\frac{1}{-v}\right)P_3\left[i_{0f,sinu} + \frac{P_3 - P_4}{P_3}\Delta i_{0f,dist}\right] \end{aligned} \tag{22}$$

$$\begin{aligned} i_{0L} &= \left(\frac{1}{-v}\right)[P_5 i_{0f} - P_6\Delta i_{0f,dist}] \\ &= \left(\frac{1}{-v}\right)P_5\left[i_{0f,sinu} + \frac{P_5 - P_6}{P_5}\Delta i_{0f,dist}\right] \end{aligned} \tag{23}$$

where, $P_1 \sim P_6$ are all positive constants and expressed as: $P_1 = -v + \frac{C_{0n}}{C_{0\Sigma}}$, $P_2 = \frac{C_{0n}}{C_{0\Sigma}} \cdot \frac{1-k^2}{1 - k^2\omega_0^2 LC_{0\Sigma}}$, $P_3 = \frac{C_{0i}}{C_{0\Sigma}}$, $P_4 = \frac{C_{0i}}{C_{0\Sigma}} \cdot \frac{1-k^2}{1 - k^2\omega_0^2 LC_{0\Sigma}}$, $P_5 = 1 - v$, and $P_6 = \frac{1-k^2}{1 - k^2\omega_0^2 LC_{0\Sigma}}$, respectively.

Obviously, (21)~(23) are all in the form of (5). As $P_3 - P_4 < 0$, $i_{0i(i\neq n)}$ is with the negative superposition, which is opposite to that of $i_{0f}$. According to (20), $P_6 = \frac{1-k^2}{1 - k^2\omega_0^2 LC_{0\Sigma}} > \frac{1}{\omega_0^2 LC_{0\Sigma}} = 1 - v = P_5$, so $P_5 - P_6 < 0$. Therefore, the superposition characteristic of $i_{0L}$ is also opposite to that of $i_{0f}$. To identify the faulty feeder, the superposition characteristic of $i_{0n}$ would better be the positive superposition, i.e., the same as that of $i_{0f}$. To meet this purpose, the $P_1 - P_2$ in (24) should be larger than 0.

$$\begin{aligned} P_1 - P_2 &= -v + \frac{C_{0n}}{C_{0\Sigma}} - \frac{C_{0n}}{C_{0\Sigma}} \cdot \frac{1-k^2}{1 - k^2\omega_0^2 LC_{0\Sigma}} \\ &= (-v)\frac{1 - k^2\omega_0^2 L(C_{0\Sigma} - C_{0n})}{1 - k^2\omega_0^2 LC_{0\Sigma}} \end{aligned} \tag{24}$$

To make $P_1 - P_2 > 0$, it should be satisfied that $1 - k^2\omega_0^2 L(C_{0\Sigma} - C_{0n}) < 0$, i.e., $\frac{C_{0n}}{C_{0\Sigma}} < 1 - \frac{1}{k^2\omega_0^2 LC_{0\Sigma}} = 1 - \frac{1-v}{k^2}$. When $v \geq -0.1$ and $k \geq 2$, it should be satisfied that $1 - \frac{1-v}{k^2} \geq 0.725$. As a result, only when $\frac{C_{0n}}{C_{0\Sigma}} < 0.725$ can make $P_1 - P_2 > 0$ always valid on the condition of $v \geq -0.1$. The larger the $\frac{C_{0n}}{C_{0\Sigma}}$, the smaller the $P_1 - P_2$ and the weaker the distortion of $i_{0n}$. In most cases, HIFs happen in overhead lines instead of underground cables, and there are more than one feeders connected to a substation, so $\frac{C_{0n}}{C_{0\Sigma}} \ll 0.725$.

In summary, at the RGN network, the ZSCs at healthy and faulty feeders are with the inverse superposition characteristics. It makes significant differences between their features. For example, phases of the same harmonic components are completely the opposite, and the current distortions also present notably different shapes, as illustrated in Fig.5(h).

*3) Verification:*

A real-world HIF experimented in the 10kV system and grounded to wet soil, is used to verify the above conclusion that is based on (21)~(23). Fig.8 (a) shows the ZSCs measured at M1~M5 and Mf. Positions of these devices have been introduced in Section II.B. The verification is conducted as follows:

Step 1: Resolve $i_{0f}$ (i.e., $i_{0Mf}$ for this case) into a sinusoidal component and a distortional component, as in Fig.8 (b).

Step 2: Use FFT to calculate the peak amplitudes of sinusoidal components for each ZSC, i.e.,: $A_{Mi(i=1,2,\ldots,5,f)}$.

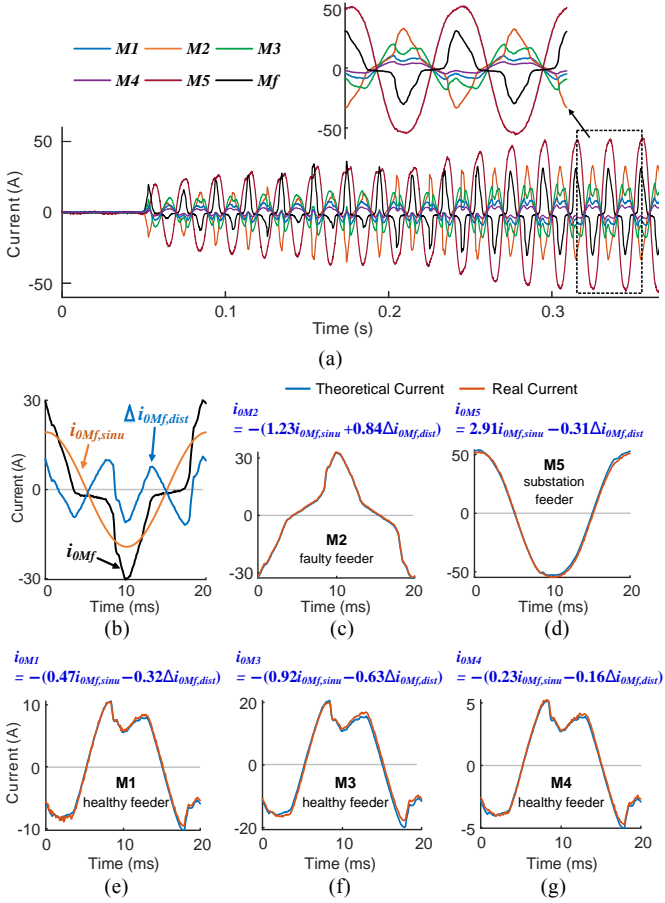

Fig.8 Verification of the theoretically calculated ZSCs by a real-world HIF. (a) ZSCs of a real HIF; (b) sinusoidal and distortional components of the ZSC at the fault point; (c)~(g) comparison between the theoretical and real ZSCs at the faulty feeder, substation feeder, and healthy feeders, respectively.

Step 3: Use these amplitudes to calculate $v$ according to (14). As explained by (20), simplify $\frac{1-k^2}{1-k^2\omega_0^2 LC_{0\Sigma}}$ as a constant, which equals $\left(\frac{1}{\omega_0^2 LC_{0\Sigma}} + \frac{3}{4\omega_0^2 LC_{0\Sigma}-1}\right)/2$.

Step 4: Bring the parameters calculated in Step 3 back into (21)~(23), and calculate the theoretical ZSCs of each feeder.

The theoretical waveforms and the real-world waveforms in Fig.8 (a) are compared in Fig.8 (c)~(g), respectively for each feeder. The expressions of (21)~(23) are calculated and presented above the corresponding figure. As is shown, the theoretical curves are perfectly consistent with the real ones, which validate the correctness of the deducing in this section.

## IV. CALIBRATION AFTER CONSIDERING DAMPING RATIO

### A. Existence of Resistive Components

At practical networks, there always exist resistive components in $i_{0f}$, which are mainly caused by the resistance of the Petersen coil and phase-to-earth resistances of transmission lines. Therefore, the equivalent circuit in Fig.7 can be transformed as Fig.9.

In some cases, resistive components are so small, so that the conclusions in the previous section are still valid, like the case in Fig.8. However, as the decrease of $v$, $i_{0L}$ is compensated to a small value, and $i_{0R}$ cannot be neglected anymore. For example, Fig.10 (a) shows the ZSCs of another real-world HIF, which is grounded to wet asphalt. Affected by $i_{0R}$, the ZSC of faulty feeder shows obvious phase deviation compared to those of healthy feeders. Due to $i_{0R_{0i}}$, phases of ZSCs at the healthy feeders are also a bit different from each other.

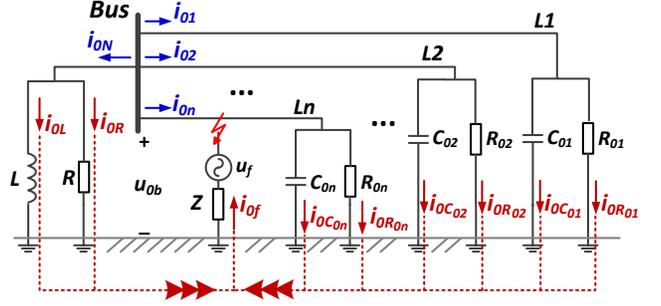

Fig.9 Equivalent zero-sequence circuit after considering damping ratio.

The blue curves in Fig.10 (c)~(g) represent the theoretical ZSCs calculated by (21)~(23). Obviously, both the phases and distortion shapes are not consistent with the real ones at all. Therefore, calibrations are required.

### B. Expressions of ZSCs at Different Feeders

The damping ratio, denoted as $d$, is defined as:

$$d = \frac{A_{MR} + \sum_{i=1}^{n} A_{MR_{0i}}}{\sum_{i=1}^{n} A_{MC_{0i}}} = \frac{1/R + \sum_{i=1}^{n} 1/R_{0i}}{\omega_0 C_{0\Sigma}} = \frac{1}{\omega_0 R_\Sigma C_{0\Sigma}} \quad (25)$$

where, $\frac{1}{R_\Sigma} = \frac{1}{R} + \sum_{i=1}^{n}\frac{1}{R_{0i}}$; $A_{MR}$ and $A_{MR_{0i}}$ represent the peak amplitudes of $i_{0R,sinu}$ and $i_{0R_{0i},sinu}$, respectively. For an RGN network, $d$ and $v$ are two significant parameters to reflect the compensation ability for the fault current.

The deducing process is similar to Section III. However, the difference is that $i_{0R}$ and $i_{0R_{0i}}$ are included in (7)~(11). Define $\overrightarrow{E_{x,sinu}} = A_0\sin(\omega_0 t + \varphi_0)$ and $\overrightarrow{E_{x,dist}^{(H,k)}} = A_{H,k}\sin(k\omega_0 t + \varphi_{H,k})$, then the $i_{0f}$ in (5) can be expressed as:

$$i_{0f} = i_{0f,sinu} + \sum_{k=2}^{m}\Delta i_{0f,dist}^{(H,k)} = \overrightarrow{E_{x,sinu}} + \sum_{k=2}^{m}\overrightarrow{E_{x,dist}^{(H,k)}} \quad (26)$$

Afterward, like (12), establish the NHL equations for sinusoidal and distortional components, respectively:

$$\begin{cases} i_{0f,sinu} = i_{0L,sinu} + \frac{L}{R_\Sigma}\frac{di_{0L,sinu}}{dt} + LC_{0\Sigma}\frac{d^2 i_{0L,sinu}}{dt^2} \\ \qquad = A_0\sin(\omega_0 t + \varphi_0) = \overrightarrow{E_{x,sinu}} \\ \Delta i_{0f,dist}^{(H,k)} = \Delta i_{0L,dist}^{(H,k)} + \frac{L}{R_\Sigma}\frac{d\Delta i_{0L,dist}^{(H,k)}}{dt} + LC_{0\Sigma}\frac{d^2\Delta i_{0L,dist}^{(H,k)}}{dt^2} \\ \qquad = A_{H,k}\sin(k\omega_0 t + \varphi_{H,k}) = \overrightarrow{E_{x,dist}^{(H,k)}} \end{cases} \quad (27)$$

Solve the NHL equations and obtain the two components of $i_{0L}$, $i_{0R}$, $i_{0C_{0i}}$, and $i_{0R_{0i}}$, respectively. Then, the ZSCs at the faulty feeder ($i_{0n}=-i_{0f}+i_{0C_{0n}}+i_{0R_{0n}}$), the healthy feeders ($i_{0i}=i_{0C_{0i}}+i_{0R_{0i}}$, $i\neq n$), and the substation feeder ($i_{0N}=i_{0L}+i_{0R}$) are respectively expressed as (28)~(30), where $v_{H,k} = 1 - \frac{1-v}{k^2}$, $d_{H,k} = \frac{d}{k}$, $r_{R_{0i}} = \frac{R_\Sigma}{R_{0i}}$, $r_R = \frac{R_\Sigma}{R}$ and $c_i = \frac{C_{0i}}{C_{0\Sigma}}$.

$$i_{0n} = \sqrt{\frac{(c_n-v)^2+d^2(r_{R_{0n}}-1)^2}{v^2+d^2}}\overrightarrow{E_{x,sinu}} \cdot e^{j\Delta\theta_{sinu}} + \sum_{k=2}^{m}\sqrt{\frac{(c_n-v_{H,k})^2+d_{H,k}^2(r_{R_{0n}}-1)^2}{v_{H,k}^2+d_{H,k}^2}}\overrightarrow{E_{x,dist}^{(H,k)}} \cdot e^{j\Delta\theta_{dist}^{(H,k)}} \quad (28)$$

$$\Delta\theta_{sinu} = \pi + \arctan\left[\frac{dc_n - r_{R_{0n}}dv}{c_nv - v^2 - (1-r_{R_{0n}})d^2}\right], \quad \Delta\theta_{dist}^{H,k} = \begin{cases} \arctan\left[\frac{d_{H,k}c_n - r_{R_{0n}}d_{H,k}v_{H,k}}{c_nv_{H,k} - v_{H,k}^2 - (1-r_{R_{0n}})d_{H,k}^2}\right], & c_n > v_{H,k} + (1-r_{R_{0n}})d_{H,k}^2/v_{H,k} \\ \pi + \arctan\left[\frac{d_{H,k}c_n - r_{R_{0n}}d_{H,k}v_{H,k}}{c_nv_{H,k} - v_{H,k}^2 - (1-r_{R_{0n}})d_{H,k}^2}\right], & c_n \leq v_{H,k} + (1-r_{R_{0n}})d_{H,k}^2/v_{H,k} \end{cases}$$

$$i_{0i(i\neq n)} = \sqrt{\frac{r_{R_{0i}}^2d^2+c_i^2}{v^2+d^2}}\overrightarrow{E_{x,sinu}} \cdot e^{j\Delta\theta_{sinu}} + \sum_{k=2}^{m}\sqrt{\frac{r_{R_{0i}}^2d_{H,k}^2+c_i^2}{v_{H,k}^2+d_{H,k}^2}}\overrightarrow{E_{x,dist}^{(H,k)}} \cdot e^{j\Delta\theta_{dist}^{(H,k)}} \quad \Delta\theta_{sinu} = \begin{cases} \arctan\left(\frac{c_id - r_{R_{0i}}dv}{c_iv + r_{R_{0i}}d^2}\right), & c_iv + r_{R_{0i}}d^2 > 0 \\ \pi + \arctan\left(\frac{c_id - r_{R_{0i}}dv}{c_iv + r_{R_{0i}}d^2}\right), & c_iv + r_{R_{0i}}d^2 \leq 0 \end{cases} \quad (29)$$

$$\Delta\theta_{dist}^{H,k} = \arctan\left(\frac{c_id_{H,k} - r_{R_{0i}}d_{H,k}v_{H,k}}{c_iv_{H,k} + r_{R_{0i}}d_{H,k}^2}\right)$$

$$i_{0N} = -\frac{\sqrt{(v^2+d^2)[(1-v)^2+r_R^2d^2]}}{v^2+d^2}\overrightarrow{E_{x,sinu}} \cdot e^{j\Delta\theta_{sinu}} - \sum_{k=2}^{m}\frac{\sqrt{(v_{H,k}^2+d_{H,k}^2)[(1-v_{H,k})^2+r_R^2d_{H,k}^2]}}{v_{H,k}^2+d_{H,k}^2}\overrightarrow{E_{x,dist}^{(H,k)}} \cdot e^{j\Delta\theta_{dist}^{(H,k)}} \quad (30)$$

$$\Delta\theta_{sinu} = \pi + \arctan\left[\frac{d(1-v) + r_Rdv}{v(1-v) - r_Rd^2}\right], \quad \Delta\theta_{dist}^{H,k} = \begin{cases} \arctan\left[\frac{d_{H,k}(1-v_{H,k}) + r_Rd_{H,k}v_{H,k}}{v_{H,k}(1-v_{H,k}) - r_Rd_{H,k}^2}\right], & v_{H,k}(1-v_{H,k}) > r_Rd_{H,k}^2 \\ \pi + \arctan\left[\frac{d_{H,k}(1-v_{H,k}) + r_Rd_{H,k}v_{H,k}}{v_{H,k}(1-v_{H,k}) - r_Rd_{H,k}^2}\right], & v_{H,k}(1-v_{H,k}) \leq r_Rd_{H,k}^2 \end{cases}$$

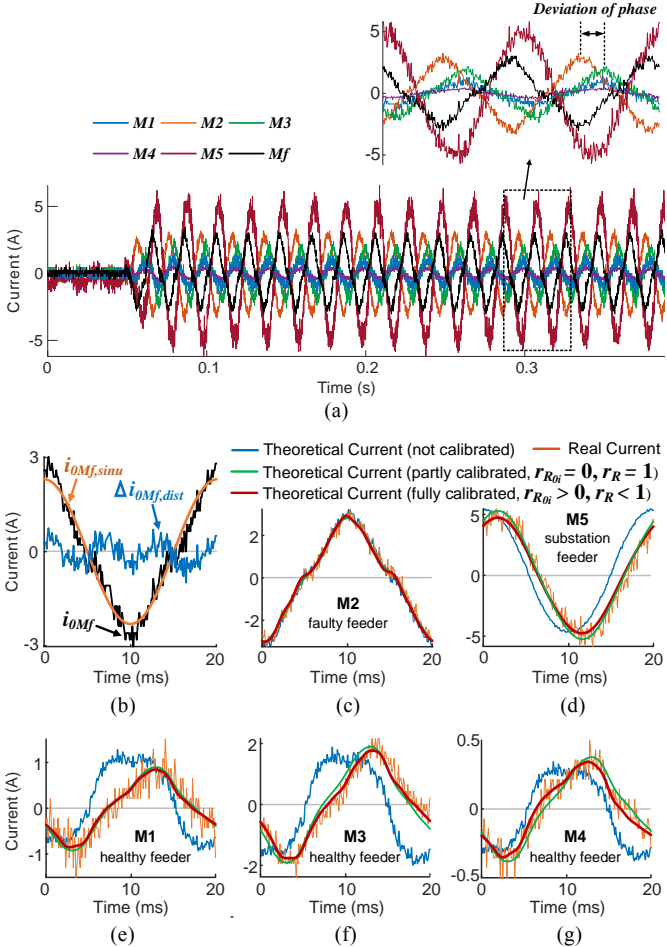

Fig.10 Comparison between real-world ZSCs (the HIF grounded to wet asphalt at RGN network) and the theoretical ZSCs without/with calibration.

According to (26) and (28)~(30), after resolving $i_{0f}$ into $\overrightarrow{E_{x,sinu}}$ and $\sum_{k=2}^{m}\overrightarrow{E_{x,dist}^{(H,k)}}$, the ZSCs at different feeders can be achieved by counterclockwise rotating $\overrightarrow{E_{x,sinu}}$ and each $\overrightarrow{E_{x,dist}^{(H,k)}}$ by different angles, multiplying them by different multipliers, and finally, linearly adding all the components up. The expressions also show that various current components of feeders all have specific relationships with those of the fault current $i_{0f}$. Although the phases and amplitudes of $i_{0f}$ vary as the impedances of feeders and fault ($Z$ in Fig.7), these relationships will not be affected. This conclusion is crucial to guarantee the universality of feeder identification method.

*C. Verification*

The real-world HIF in Fig.10 (a) is used to verify the correctness of (28)~(30), the coefficients of which are calculated as follows:

Step 1: Resolve $i_{0f}$ (i.e., $i_{0Mf}$ for this case) into $i_{0f,sinu}$ and $\Delta i_{0f,dist}$ as in Fig.10 (b), which are represented as $\overrightarrow{E_{x,sinu}}$ and $\sum_{k=2}^{m}\overrightarrow{E_{x,dist}^{(H,k)}}$, respectively.

Step 2: Fig.11 shows a phasor diagram when considering the damping ratio. Use FFT to calculate the phases of $u_{0b,sinu}$, $i_{0i,sinu(i\neq n)}$, and $i_{0N,sinu}$ as $\varphi_{0b}$, $\varphi_{0i}$, and $\varphi_{0N}$, respectively. Then, the phases of $i_{0C_{0i},sinu}$ and $i_{0L,sinu}$ can be estimated as $\varphi_{0b} + \frac{\pi}{2}$ and $\varphi_{0b} - \frac{\pi}{2}$, respectively.

Step 3: Calculate the peak amplitudes of $i_{0i,sinu(i\neq n)}$ and $i_{0N,sinu}$ by FFT. Then, with all phases are known, the peak amplitudes of $i_{0C_{0i},sinu}$, $i_{0R_{0i},sinu}$, $i_{0L,sinu}$, and $i_{0R,sinu}$ can be calculated as $A_{MC_{0i}}$, $A_{MR_{0i}}$, $A_{ML}$, and $A_{MR}$, respectively.

Step 4: With the peak amplitudes in Step 3, calculate $v$ and $d$ according to (14) and (25), as well as $r_{R_{0i}}$, $r_R$ and $c_i$. Finally, the theoretical ZSCs at different feeders are calculated by (28)~(30) and presented in Fig.10 (c)~(g). Only the harmonics below the 11th order ($k \leq 11$) are considered.

In Fig.10 (c)~(g), green curves are the ZSCs only considering the resistive component of Petersen coil ($r_{R_{0i}}=0, r_R=1$), whereas the red curves also consider the phase-to-earth resistances ($r_{R_{0i}}>0, r_R<1$). Compared to the waveforms without calibration (colored blue), the calibrated curves show much more satisfying fitting results. Besides, the accuracy can be further improved after considering phase-to-earth resistances.

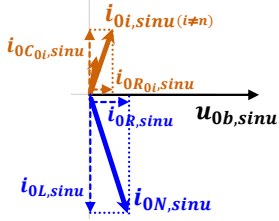

Fig.11 Phasor diagram of zero-sequence signals considering damping ratio.

## V. APPLICATION IN FEEDER IDENTIFICATION

### A. A Classic Method and Its Improvement

Many algorithms detect HIFs based on the 3rd harmonic component as it is the dominant cause of current nonlinearity. For example, a traditional HIFAS system built up by Nordon Technologies [3] and T. Cui's work [7] both believe that the phase difference between the fundamental current and the 3rd harmonic current (denoted as $\Delta\varphi_{I1-I3}$) is around 180° (e.g., 180°±40°) at the fault area. This well-known fault information reveals an original nonlinearity feature of HIF, inspiring many further algorithms based on distortions [12], [22] and time-frequency distributions [1], [11]. These approaches are effective to identify the feeder of HIF when the damping ratio is small. Take the HIF in Fig.8 as an example, the average $\Delta\varphi_{I1-I3}$ of the faulty feeder (M2) is 185.8°, while those of healthy feeders (M1, M3, and M4) are 1.4°, -5.2°, and 7.6°, respectively.

However, with the theories introduced in this paper, we can make a confident judgment that the above algorithms cannot work under many fault conditions, for the following two factors:

Factor 1): the diversity of distortions introduced in Section II, especially the phenomenon of 'distortion offset', could make the above methods invalid. For example, $\Delta\varphi_{I1-I3}$ is about 267.8° and 255.6° for the HIFs in Fig.5 (f) and (g), respectively.

Factor 2), the existence of the damping ratio make it unreliable to distinguish between the faulty feeder and healthy feeders only by $\Delta\varphi_{I1-I3}$. Denote the $\Delta\varphi_{I1-I3}$ at the healthy feeder $i$ as $\Delta\varphi_{0i,I1-I3}$, and at the fault point as $\Delta\varphi_{0f,I1-I3}$. To simplify analyses, phase-to-earth resistances of feeders are neglected ($r_{R_{0i}}=0$). Then, according to (29), $\Delta\varphi_{0i,I1-I3}$ is expressed as:

$$\Delta\varphi_{0i,I1-I3} = \Delta\varphi_{0f,I1-I3} + 3\Delta\theta_{sinu} - \Delta\theta_{dist}^{(H,3)}$$
$$= \Delta\varphi_{0f,I1-I3} + 3\left[180° + arctan\left(\frac{d}{v}\right)\right] \quad (31)$$
$$- arctan\left(\frac{3d}{8+v}\right)$$

When considering *Factor 1)*, we can set $\Delta\varphi_{0f,I1-I3} \in [180°-40°, 267.8°+40°]$. In (31), it can be calculated that $3\left[180° + arctan\left(\frac{d}{v}\right)\right] - arctan\left(\frac{3d}{8+v}\right) \in [-100°, 180°)$, when $v \in [-0.1, 0)$ and $d \in [0, 0.5]$. Then, $\Delta\varphi_{0i,I1-I3}$ should be in the range of $[40°, 487.8°)$ It means the $\Delta\varphi_{I1-I3}$ of healthy feeder could be any angle and even equal to that of the faulty feeder. For the case in Fig.10, the average $\Delta\varphi_{I1-I3}$ of the faulty feeder (M2) is 196.6°, while those of healthy feeders (M1, M3, and M4) are 197.1°, 188.4°, and 238.9°, respectively. As a result, just $\Delta\varphi_{I1-I3}$ cannot guarantee the reliability of feeder identification in RGN system. Improvements are needed.

Based on the theoretical analysis in Section IV, this paper is meant to provide a novel faulty feeder identification method to keep the reliability under as many conditions as possible. In terms of harmonics, the 3rd harmonic is still the most encouraged as its content is the largest.

The key criterion of the proposed method is as follows. Suppose that $\varphi_{0i,dist}^{(H,3)}$ represents the phase of 3rd harmonic current in feeder $i$ ($i=1,2,…,n$), and it can be represented by summing the phase of fault current $\varphi_{0f,dist}^{(H,3)}$ and a rotated angle $\Delta\theta_{0i,dist}^{(H,3)}$, according to (28) and (29):

$$\varphi_{0i,dist}^{(H,3)} = \varphi_{0f,dist}^{(H,3)} + \Delta\theta_{0i,dist}^{(H,3)} \quad (32)$$

where, $\Delta\theta_{0i,dist}^{(H,3)}$ is the $\Delta\theta_{dist}^{(H,3)}$ in (28) and (29).

For the feeder $n$, make a judgment that whether its phase difference from another feeder $i$ ($i\neq n$) satisfies the requirement in (33). If it is satisfied for all feeders, the feeder $n$ will be regarded as the faulty feeder.

$$\begin{aligned} Indicator &= \left|\varphi_{0n,dist}^{(H,3)} - \varphi_{0i,dist}^{(H,3)} - 180°\right| \\ &= \left|\Delta\theta_{0n,dist}^{(H,3)} - \Delta\theta_{0i,dist}^{(H,3)} - 180°\right| \\ &\leq Thr \end{aligned} \quad (33)$$

In practical applications, indicators can be obtained by directly calculating $\varphi_{0n,dist}^{(H,3)}$ and $\varphi_{0i,dist}^{(H,3)}$ with the Fourier transformer. To better understand the performance of the indicator, the expressions of $\Delta\theta_{0n,dist}^{H,3}$ in (28) and $\Delta\theta_{0i,dist}^{H,3}$ in (29) are brought into (33). Then, the curves of indicators versus $v$, $d$, and $c_n$ are shown in Fig.12. $Thr$ is set as 40° in this paper.

A major advantage of the proposed indicator is that it is not affected by the aforementioned *Factor 1)*. The 'distortion offset' only changes the phase of fault current, i.e., the $\varphi_{0f,dist}^{(H,3)}$ in (32), but will not affect the phase differences in (33). Besides, for Factor 2), when $c_n$ is smaller than 0.4, the 'indicator' is effective for all $v$ and $d$. Although the effective area will be narrowed as the increase of $c_n$, the proposed indicator is effective under most common conditions.

To intuitively present the strengths of the method, the comparison is carried out. Fig.13 shows the effective area of the classic method mentioned at the start of this section. If the threshold is also set as 40°, the criteria should have been $\left|\Delta\varphi_{0n,I1-I3}-180°\right|\leq 40°$ for the faulty feeder and $|\Delta\varphi_{0i,I1-I3\ (i\neq n)}-180°|>40°$ for healthy feeders. To simplify the problem, the $\Delta\varphi_{0f,I1-I3}$ in (31) is assumed as 180°, which means the Factor 1) is neglected. Then, the criteria will be the equations exhibited in Fig.13, where the arrows represent the directions that the criteria are satisfied. Comparing Fig.12 with Fig.13, the proposed indicator presents conspicuous improvements.

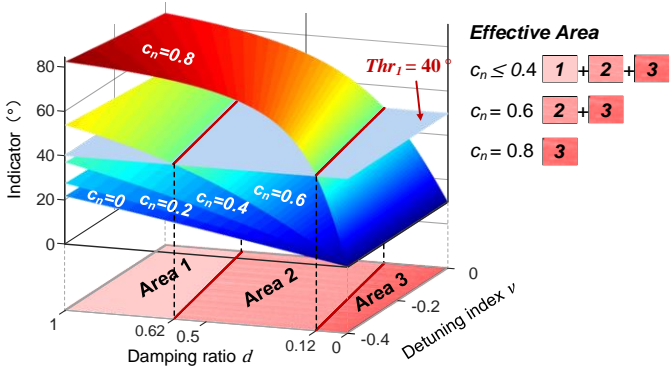

Fig.12 Effective area of the proposed 'indicator'.

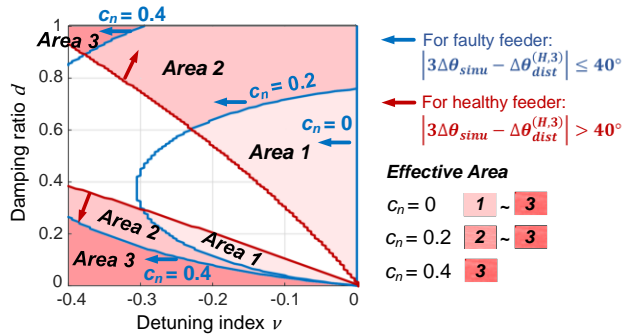

Fig.13 The effective area of a widely used traditional approach. *Factor 1)* is not considered and $\Delta\varphi_{0f,I1-I3}$ is simplified as a constant 180°.

### B. Case Study and Robustness Assessment

With the proposed method, Fig.14 presents the feeder identification results of four HIFs, which are experimented in the real-world 10kV distribution system. According to Fig.4, there are four feeders in the system and the zero-sequence currents are respectively measured by M1~M4 at the start of feeders. For the resonant three-wire system, zero-sequence currents are close to zero on the normal conditions due to the isolation of ungrounded load transformers. It makes identifying the feeder of HIF that is only with several amperes become possible.

In the four HIF experiments of Fig.14(a)~(d), line conductors are grounded to different surfaces. The fault impedances presented in the title of Fig.14 are estimated according to the fault voltage and current. There are five graphs in each figure. The first graph presents the zero-sequence currents measured at the start of four feeders. The other four graphs respectively represent the indicators of each feeder. For example, the indicators of current measured by M1 in Fig.14(a) include the phase differences (the indicator in (33)) from that of M2, M3, and M4, respectively. As a result, there are three curves in each of the four graphs. The indicators are all calculated twice per cycle. Synchronism of measurement must be guaranteed when calculating the phase differences.

The damping ratios $d$ of the HIFs in Fig.14(a) and (b) are negligible, while those in Fig.14(c) and (d) are about 0.4. This difference can be noticed by observing the different phase deviations of the faulty feeder current from those of healthy feeders. According to Fig.14(a)~(c), for the faulty feeder, the indicators can all stably keep within the range of $Thr$. For healthy feeders, indicators are within the $Thr$ only when calculating their phase differences from the faulty feeder. The indicators are always near 180° when calculating the phase differences between healthy feeders. That is, the indicator curves of the faulty feeder are all within $Thr$, while there is only one indicator curve within $Thr$ for the healthy feeder, as shown in Fig.14(a)~(c). The HIF in Fig.14(d) is a special case, the fault impedance of which is so high at the beginning of fault that the fault current is smaller than the measuring scale. As the development of HIF, the ground surface is melted by arcs, and the fault impedance decreases, so the indicators start concentrating within $Thr$ at about 0.65s after fault happens. The four cases verify the effectiveness of the proposed method, even under some extremely challenging situations.

Robustness to normal harmonics, like load harmonics, is significant for a harmonic-based method. The robustness requires that a method should not make misjudgments no matter under normal or fault conditions:

*1) Normal condition*. Assume that the harmonic load is at the end of feeder n. If there is no fault happening, the equivalent zero-sequence network of the resonant three-wire system is that in Fig.15. As is shown, the current path to harmonic load is regarded as open-circuited due to the isolation of the ungrounded transformer, which means $i_{0Ldn}=0$. However, the harmonic components can still be induced to the high-voltage side as the terminal voltage $u_{0Ldn}$, and reflected by the unbalanced voltage $u_{unb}$. With the filter of delta wiring of transformer, the induced 3rd harmonic voltage is so small compared to the fundamental component that can be neglected. Besides, feeder identification is usually triggered after a HIF is successfully detected. If the HIF detection algorithm is reliable under the operation of normal harmonic loads, the proposed method won't be started at all.

*2) Fault condition*. Assume the harmonic load is still connected at the end of feeder *n* and the HIF happens at the same feeder as well. Due to the isolation of the load transformer, the equivalent zero-sequence network is similar to that in Fig.9, That is, there is also no path for zero-sequence current to the harmonic load. The only difference from Fig.9 is that the virtual voltage $u_f$ will contain the harmonic components generated by the harmonic load. It does affect the phases and amplitudes of the harmonic components of $i_{0f}$. However, the expressions of currents in different feeders are deduced by taking $i_{0f}$ as the reference, according to the basic equation in (27) and final expressions in (28)~(30). As a result, although the $\varphi_{0f,dist}^{(H,3)}$ in (32) is changed by the harmonic component of $u_f$, the indicators determined by phase differences of $\Delta\theta_{0i,dist}^{(H,3)}$ in (33) possess the complete robustness to it.

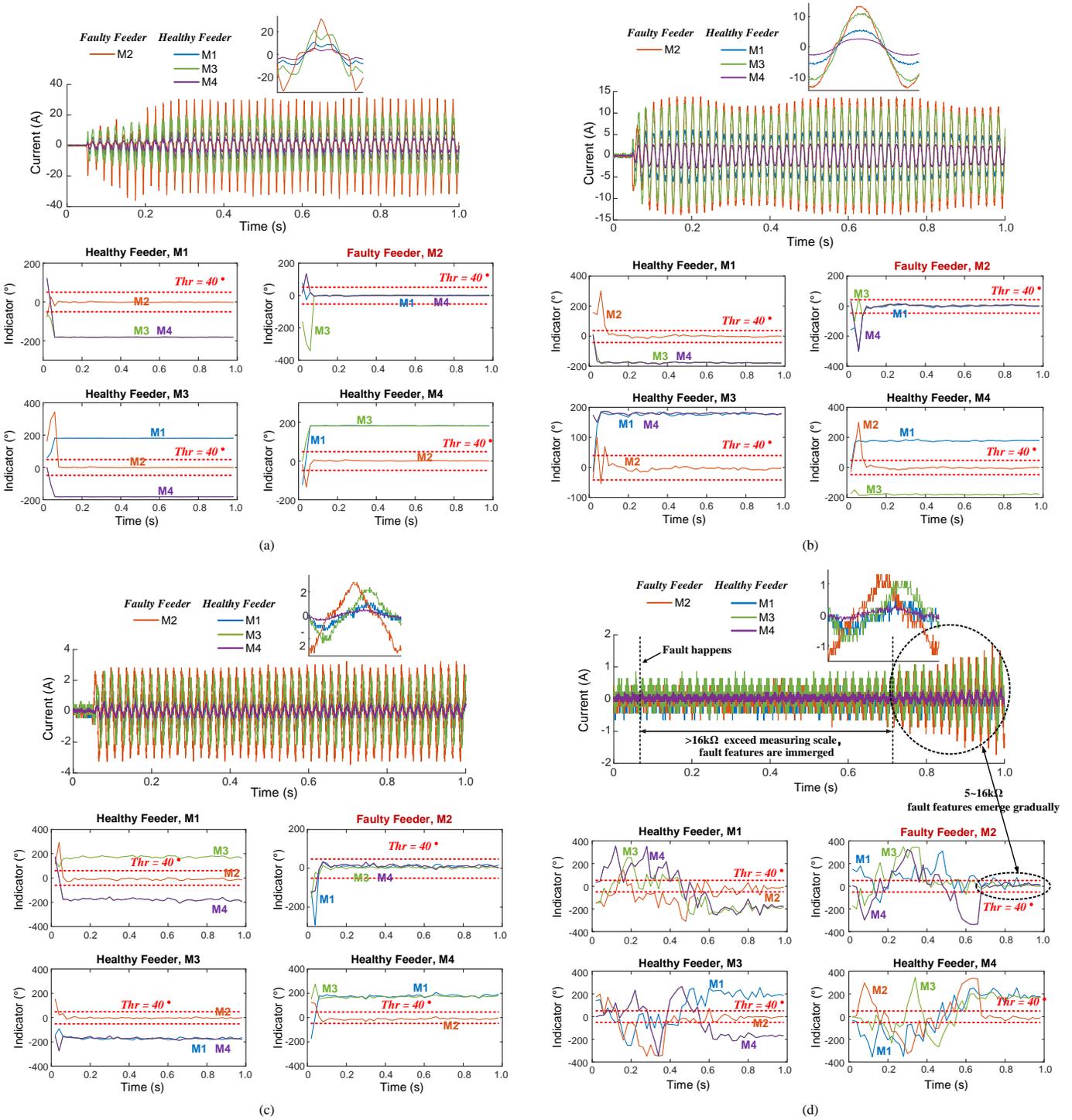

Fig.14 Feeder identification results of real-world HIFs grounded with different surfaces: (a) wet soil (300Ω~1kΩ), (b) wet cement (600Ω~800Ω), (c) dry asphalt (3kΩ), (d) dry cement (>5kΩ).

A 4-feeder (L1~L4) resonant system is modeled in PSCAD, and its topology is just like the real-world system in Fig.4. A harmonic load (arc furnace) is connected at the end of L4, which switches in at 0.1s. The 3rd harmonic current of load accounts for 12% of the fundamental component. At the same feeder, a 2 kΩ HIF is switched in at 0.3 s. The HIF is simulated by the DIST-C model proposed in [19], where the NO.13 parameter combination in TABLE II is selected to simulate the nonlinearity. Fig.16(a) shows phase voltages and currents measured at the start of L4, where the load harmonics are conspicuous in B-phase and C-phase currents. As shown in Fig.16(b), the increase of zero-sequence signals caused by the switch-in of the harmonic load is much smaller than that caused by HIF, as the current can only flow through the large phase-to-earth impedances. According to the indicators in Fig.16(c), the proposed indicators will not make mistakes under the operation of normal harmonic load (0.1~0.3s). Meanwhile, under the interference of harmonic load, the feeder of HIF can still be correctly identified (after 0.3s). The robustness of the proposed method is thereby verified.

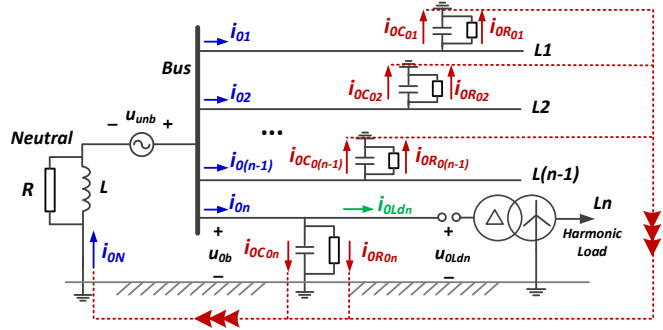

Fig.15 Equivalent zero-sequence circuit under normal condition, considering the connection of harmonic load.

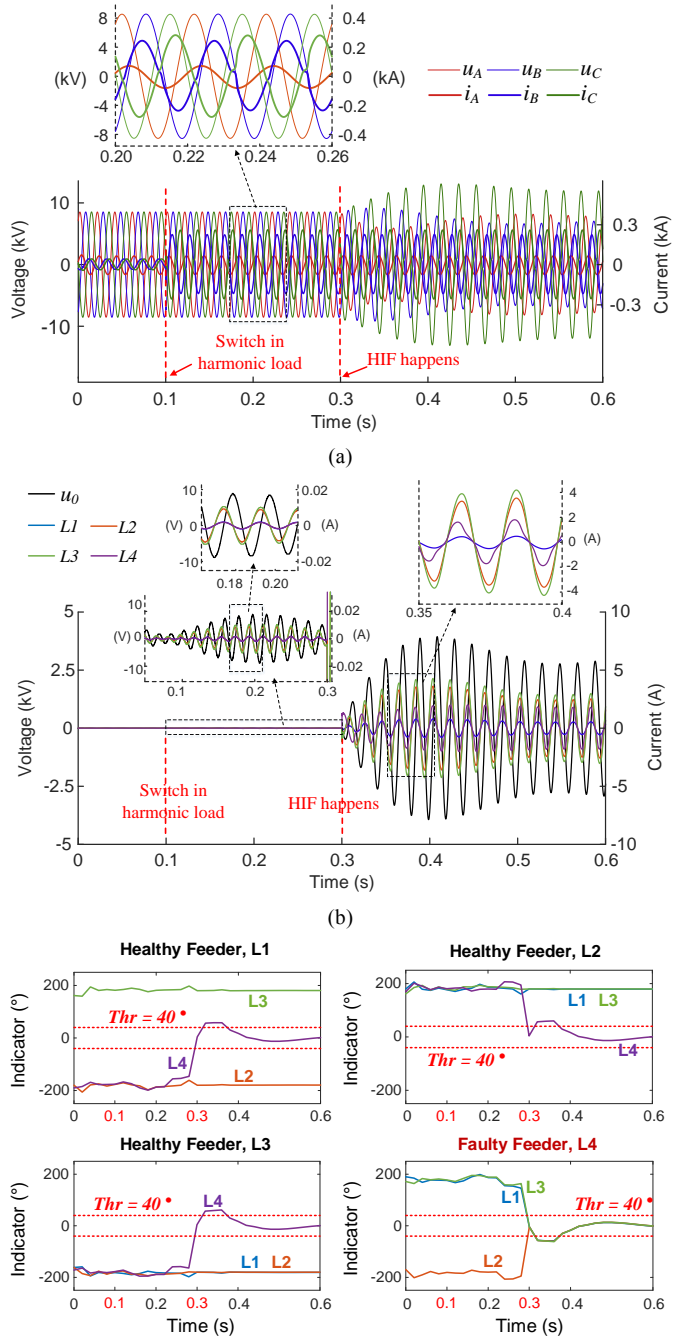

Fig.16 Waveforms and results when considering the harmonic load: (a) phase voltages and currents of the faulty feeder, L4; (b) zero-sequence currents of different feeders and the zero-sequence voltage; (c) indicators of different feeders.

## VI. CONCLUSION

This paper provides a theoretical basis to reveal the essential features of HIFs at different feeders, and help find out the most suitable fault information for feeder identification.

1) Based on the energy balance theory, the diversity of nonlinear distortion is explained and confirmed by real HIFs;

2) Characteristics of distortions and phase relationships are theoretically deduced. Both the effects of the detuning index and damping ratio at practical networks are considered. Conclusions are all verified by real HIF. The theories in this paper help understand the relationship between different ZSC components, catch the essential differences between feeders, and provide solid guidance to more reliable approaches;

3) A specific phase relationship based on the 3$^{rd}$ harmonic ZSC is formed as a universally reliable indicator to identify the feeder of HIF. Real-world cases of HIFs verify the effectiveness and advantages.

REFERENCE

[1] A. Ghaderi, et al, "High-Impedance Fault Detection in the Distribution network Using the Time-Frequency-Based Algorithm," *IEEE Trans. Power Del.*, vol. 30, no. 3, pp. 1260-1268, Jun. 2015.
[2] M. Mishra, and R. R. Panigrahi, "Taxonomy of high impedance fault detection algorithm," Measurement, vol. 148, 106955, 2019.
[3] Power System Relaying Committee, Working Group D15 of the IEEE Power Eng. Soc., "High impedance fault detection technology," Mar. 1996.
[4] CIGRE Study Committee B5, Report of Working Group 94, "High impedance faults," Jul. 2009.
[5] D. P. S. Gomes, et al. "The effectiveness of different sampling rates in vegetation high-impedance fault classification," *Electr. Power Syst. Res.*, vol. 174, 2019.
[6] A. Ghaderi, H.L. Ginn III and H. A. Mohammadpour. "High impedance fault detection: A review." *Elec. Power Syst. Res.,* 143: pp: 376-388, 2017.
[7] T. Cui, X. Dong, Z. Bo and S. Richards, "Integrated scheme for high impedance fault detection in MV distribution system," *2008 IEEE/PES Transmission and Distribution Conference and Exposition*: Latin America, Bogota, 2008, pp. 1-6.
[8] D. P. S. Gomes, C. Ozansoy and A. Ulhaq, "High-Sensitivity Vegetation High-Impedance Fault Detection Based on Signal's High-Frequency Contents," *IEEE Trans. Power Del.*, vol. 33, no. 3, pp. 1398-1407, Jun. 2018.
[9] Y. Xue, et al., "Resonance Analysis and Faulty Feeder Identification of High-Impedance Faults in a Resonant Grounding System," *IEEE Power Del.*, vol. 32, no. 3, pp. 1545-1555, June 2017.
[10] A. Nikander, and P. Jarventausta, "Identification of High-Impedance Earth Faults in Neutral Isolated or Compensated MV Networks," *IEEE Trans. Power Del.*, vol. 32, no. 3, pp. 1187-1195, Jun. 2017.
[11] M. Wei, et al., "High Impedance Arc Fault Detection Based on the Harmonic Randomness and Waveform Distortion in the Distribution System," *IEEE Trans. Power Del.*, vol. 35, no. 2, pp. 837-850, April 2020.
[12] M. Michalik, et. al., "High-impedance fault detection in distribution networks with use of wavelet-based algorithm," *IEEE Trans. Power Del.,* vol. 21, no. 4, Oct. 2006.
[13] X. Wang et al., "High Impedance Fault Detection Method Based on Variational Mode Decomposition and Teager–Kaiser Energy Operators for Distribution Network," *IEEE Trans. Smart Grid*, vol. 10, no. 6, pp. 6041-6054, Nov. 2019.
[14] C. Gonzalez, et al. "Directional, High-Impedance Fault Detection in Isolated Neutral Distribution Grids." *IEEE Trans. Power Del.*, vol. 33, no, 5, Oct. 2018.
[15] T. Tang, et al. "Single-phase high-impedance fault protection for low-resistance grounded distribution network," *IET Gener. Transm. Distrib.*, vol. 12, no. 10, pp. 2462-2470, 2018.
[16] K. Pandakov, H. K. Hoidalen, and S. Traetteberg, "An Additional Criterion for Faulty Feeder Selection During Ground Faults in Compensated Distribution Networks," *IEEE Trans. Power Del.*, vol. 33, no. 6, pp. 2930-2937, 2018.
[17] T. Y. Li, Y. D. Xue and B. Y. Xu, "High-impedance fault detection technology based on transient information in a resonant grounding system," *CIRED - Open Access Proceedings Journal*, vol. 2017, no. 1, pp. 1176-1179, 10 2017.

[18] Y. Xue, et al., "Practical Experiences with Faulty Feeder Identification Using Transient Signals," IET 9th International Conference on Developments in Power System Protection, Glasgow, 2008, pp. 720-723.
[19] M. Wei, et.al., "Distortion-Controllable Arc Modeling for High Impedance Arc Fault in the Distribution Network," *IEEE Trans. Power Del*, vol. 36, no. 1, pp. 52-63, Feb. 2021.
[20] Y. Xue, et al., "LC resonance mechanism analysis of fault transient for single phase earth fault in non-solidly earthed network," *2013 IEEE Power & Energy Society General Meeting*, Vancouver, BC, 2013, pp. 1-5, doi: 10.1109/PESMG.2013.6672456.
[21] P. Wang, B. Chen, H. Zhou, T. Cuihua, and B. Sun, "Fault Location in Resonant Grounded Network by Adaptive Control of Neutral-to-Earth Complex Impedance," *IEEE Trans. Power Del.,* vol. 33, no. 2, pp. 689-698, 2018.
[22] M. Wei, et.al., "Distortion-Based Detection of High Impedance Fault in Distribution Systems," *IEEE Trans. Power Del.*, 2020.